\documentclass[journal,10pt]{IEEEtran}
\usepackage[pdftex]{graphicx}
\usepackage{multirow}
\usepackage{amsmath}
\usepackage{mathtools}
\usepackage{optidef}
\usepackage{epstopdf}
 \usepackage[ruled,vlined]{algorithm2e}
\usepackage[justification=centering,font=scriptsize]{caption}
\usepackage{authblk}
\usepackage{cite}
\usepackage{cleveref}
\usepackage{latexsym}
\usepackage{amsthm}
\usepackage{balance}
\usepackage{relsize}
\usepackage[nodisplayskipstretch]{setspace}
\usepackage{color}
\usepackage{amssymb}
\usepackage{eucal}
\usepackage{url}
\usepackage{booktabs,amsfonts,dcolumn}
\newcommand\norm[1]{\left\lVert#1\right\rVert}
\usepackage{float}
\usepackage{scalefnt}
\usepackage{geometry}

\usepackage[nodisplayskipstretch]{setspace}
\usepackage{color}  
\definecolor{Mypink}{RGB}{255,0,255}
\definecolor{Myorange}{RGB}{255,102,0}
\definecolor{Mygreen}{RGB}{0,153,0}
\definecolor{Myblue}{RGB}{0,0,255}
\newcolumntype{d}[1]{D..{#1}}
\usepackage{lipsum}

\DeclareMathAlphabet\mathbfcal{OMS}{cmsy}{b}{n}
\usepackage{breqn}

\begin{document}

\title{An Effective   Spatial Modulation Based    Scheme for Indoor VLC Systems}
\renewcommand\Authfont{\fontsize{12}{14.4}\selectfont}

\author{Shimaa Naser,~Lina~Bariah,~\IEEEmembership{Senior~Member,~IEEE},~Sami~Muhaidat,~\IEEEmembership{Senior Member,~IEEE,} Mahmoud Al-Qutayri,~\IEEEmembership{Senior Member, IEEE},~and Paschalis C. Sofotasios,~\IEEEmembership{Senior Member,~IEEE}

\thanks{ This work was supported in part by Khalifa University
under Grant KU/FSU-8474000122 and Grant KU/RC1-C2PS-T2/8474000137.
\textit{Corresponding Author: Paschalis C. Sofotasios}.}

\thanks{S. Naser and L. Bariah are with the Center for Cyber-Physical Systems, Department of Electrical Engineering and Computer Science, Khalifa University, Abu Dhabi 127788, UAE, (e-mails: 100049402@ku.ac.ae; lina.bariah@ieee.org.)}

\thanks{S. Muhaidat is with the Center for Cyber-Physical Systems, Department of Electrical Engineering and Computer Science, Khalifa University, Abu Dhabi 127788, UAE, and also with the Department of Systems and Computer Engineering, Carleton University, Ottawa, ON K1S 5B6, Canada, (e-mail: muhaidat@ieee.org).}

\thanks{M. Al-Qutayri is with the Systems-on-Chip (SoC) Center, Department of Electrical Engineering and Computer Science, Khalifa University, Abu Dhabi 127788, UAE, (e-mail: mahmoud.alqutayri@ku.ac.ae).}

\thanks{P. C. Sofotasios is with the Center for Cyber-Physical Systems, Department of Electrical Engineering and Computer Science, Khalifa University, Abu Dhabi 127788,  UAE, and also with the Department of Electrical Engineering, Tampere University, Tampere 33014, Finland, (e-mail: p.sofotasios@ieee.org).}
}
\maketitle
\begin{abstract}

\textcolor{black}{We propose an enhanced spatial modulation 
(SM)-based scheme for indoor visible light communication systems. 
This scheme  enhances the achievable throughput of conventional SM schemes by transmitting higher order complex modulation symbol, which is decomposed into three different parts. 
These parts carry the amplitude, phase, and quadrant   components of the complex symbol, which are then represented by  unipolar pulse amplitude modulation (PAM) symbols. 
Superposition coding is exploited to allocate  a  fraction of the total power to each part before they are all  multiplexed and transmitted simultaneously, exploiting the entire available bandwidth. 
At the receiver, a two-step decoding process is proposed to   decode the active light emitting diode   index  before the complex    symbol is retrieved. } \textcolor{black}{
It is shown that at higher spectral efficiency values, the proposed modulation scheme outperforms conventional  SM schemes with PAM symbols  in terms of average symbol error rate (ASER),  and hence, enhancing the system throughput. 
Furthermore, since the performance of the proposed modulation scheme is sensitive to the power allocation factors, we formulated an ASER optimization problem and propose a  sub-optimal solution using successive convex programming (SCP).}  
Notably, the proposed algorithm  converges after only few iterations, whilst  the performance with the optimized power allocation coefficients outperforms both random and fixed power allocation.
\end{abstract}

\vspace{-5pt}
\section{Introduction}
Visible light communication (VLC) is envisioned to play a core role  in future wireless networks as an efficient method to complement overcrowded radio frequency (RF) systems \cite{9411734, 9226406}.
\textcolor{black}{Within this context, VLC utilizes the indoor light emitting diodes (LEDs) for both illumination and wireless  data transmission \cite{8908721}.  It is worth mentioning that  the intensity of the light  emitted from the LEDs is modulated in order to  transmit the corresponding  information signal  \cite{9199611, 8742666}. This process is usually referred to as intensity modulation (IM). On the other hand, a photo detector (PD) is utilized at the receiver   in order to retrieve the original signal, a process  known as direct detection (DD). However, VLC systems suffer from the limited modulation bandwidth of LEDs in addition to the constraints imposed by IM/DD schemes that, unlike RF communications, require the transmitted signals to be positive and real-valued \cite{9543660, 9309358}.
Hence, the implementation of spectrally efficient modulation schemes that overcome these limitations is of   paramount importance for the design of efficient high data rate VLC systems \cite{9199619, ABUMARSHOUD2021101440, 8742666, 9309358, 9117130}. } \\
\indent 
Based on the above, extensive research efforts have been devoted in order to come up with high spectral efficiency modulation techniques. 
To that end, multiple-input multiple-output (MIMO) index modulation schemes have  recently emerged as an efficient way to enhance the spectral and energy efficiency  of wireless transmission through the activation states of the building blocks of   communication systems \cite{IndexModulation}. Different investigations on spatial domain index modulation have been carried out in the context of VLC, such as space shift keying (SSK), generalized space shift keying (GSSK)  and spatial modulation (SM), wherein, the transmitting LED index is used to transmit further information \cite{Wang2018,su2021,Mesleh,Tran2019,gao2020}. In addition, multiple active SM (MA-SM) has been proposed as a generalized version of SM which is capable of enhancing the achievable  spectral efficiency by conveying more information in both spatial and signal domains \cite{Rajesh}. 
The operation of MA-SM relies on the activation of $N_{a}$ LEDs out of $N_{t}$ with multiple distinct real non-negative $M$-ary symbols transmitted from each active LED. 
Yet, it is noted that despite the undoubted advantages of these schemes, their performance   is highly affected by the incurred channel similarity associated with  the nature of VLC channels. 
\indent 

\textcolor{black}{Recent research efforts have focused  on  new techniques aiming to improve the performance of SM-based schemes in VLC systems. For instance, the authors in \cite{Rajesh} and \cite{8115222}  modified the selection of symbols to be transmitted through the active LEDs based on collaborative constellation in order to improve the power efficiency of conventional SM schemes. Furthermore, the authors in \cite{gao2020} proposed an LED grouping scheme  aiming to improve the error rate performance of SM schemes. In particular, they considered that the LEDs are separated into different groups with the purpose of alleviating the channel correlation among each group. Within the same context, in \cite{Tran2019}, the authors proposed a receiver-oriented SM scheme to efficiently mitigate the optical channel correlation across the multiple LEDs. Additionally, in \cite{9235559}, the authors proposed an augmented SM scheme that allows each transmitter to have its own identity that is embedded in the transmitted signal and can only be decoded by the indented receiver. On the contrary, non-DC biased optical orthogonal frequency division multiplexing (OFDM) was combined with SM in \cite{7495929}; however, it is rather common that OFDM suffers from the high peak-to-average power ratio, which has detrimental effects in OFDM-based VLC systems. The authors in \cite{7582437} proposed  channel-adapted SM schemes to find the optimal combinations of active LEDs under controlled inter-channel interference. Similar investigations on integrating adaptive modulation schemes with SM appeared also in \cite{8399742,9117130,8356008}. Finally, in order to enhance the error rate performance of adaptive SM schemes, the authors of \cite{9309358} proposed a flexible SM scheme that changes the modulation sizes over the LEDs and the number of active LEDs.  }

As already mentioned, the main challenge of using IM/DD schemes is the  restrictions associated with the transmitted signals. Therefore, in order to transmit complex-valued symbols,   different contributions investigated the design of multiple-LED complex modulation schemes, including quad-LED and dual-LED complex modulation. For instance, the authors in \cite{Tejaswi2016} proposed a complex modulation scheme called quad-LED complex modulation (QCM) for MIMO VLC systems. In this scheme, the spatial domain was utilized to transmit the real and imaginary parts of the complex-valued symbols. Similarly, \cite{Zhang2018}  discussed the integration of QCM and SM to further enhance the achievable transmission   rate. Furthermore, the authors in \cite{Sushanth2019}  proposed two complex modulation schemes called quadrature spatial modulation (QSM) and dual mode index modulation (DMIM) with a dual-LED complex modulator (DCM). \textcolor{black}{Nevertheless, the use of quad-LED and dual-LED complex modulations results in spectral efficiency loss when combined with SM since part of complex signal is carried over the spatial domain. Additionally the associated computational complexity with these two schemes is rather high.}

\textcolor{black}{ Based on above, the available contributions on the performance enhancement of SM have been based on two main approaches:  i) Employing multiple active LEDs with LED selection optimization and the use of complicated adaptive schemes in order to achieve a  balance  between  error rate and spectral  efficiency performance. ii) Using quad-LED and dual-LED. However, the latter approach yields an increased system complexity and limits the spectral efficiency when integrated with SM-based techniques because part of the complex symbol is conveyed through the spatial domain. Hence, in the current contribution, we aim to design an efficient higher order SM-based modulation  scheme  that can achieve an adequate balance  between  error rate performance and spectral  efficiency, while maintaining a relatively reduced  involved implementation complexity.   }
\indent 

\textcolor{black}{With this motivation, in order to improve the performance of conventional SM schemes, we propose the transmission of a superimposed signal of the amplitude, phase, and quadrant components over the signal domain and provide a formulation of the associated power allocation optimization. To the best of our knowledge, such a contribution is novel, and has not been considered previously in the open literature.} More specifically, in order to transmit a complex-valued symbol, the signal domain information is decomposed into  three  different  parts  that  are  used  to  carry  the  amplitude, phase, and quadrant components. Then,  each component is represented by a unipolar pulse-amplitude modulation (PAM) signal. Subsequently,  superposition  coding  is  exploited  to  allocate  each part  a  certain  fraction  of  the  total  power.  Then,  all  parts  are multiplexed  and  transmitted  at  the  same  time,  exploiting  the entire  available  bandwidth.
To this effect and given   that  power allocation is an essential factor that determines the overall system error performance, we also   formulate an average symbol error rate (ASER) minimization constrained optimization problem in order to find the optimum power allocation coefficients for each signal part. Then, the formulated non-convex optimization problem is effectively solved sub-optimally using successive convex programming (SCP) with trust region algorithm. 

\indent
In summary, the  main contributions of the present work are listed below:

\begin{itemize}
\item We propose a novel SM-based scheme, so called APQ-SM, that transmits complex-valued APQ symbols from the active LED.
\item We derive  closed form expressions for the ASER for both joint and two-step detection. These expressions have a relatively simple algebraic representation which renders them tractable both analytically and numerically.  
\item Capitalizing on the above, we quantify the achievable ASER performance of the proposed scheme for both joint detection and the proposed two-step detection process, which leads to the development of useful theoretical and practical insights. 
\item An efficient power allocation scheme for each part of the APQ symbol is proposed, where we formulate a constrained optimization problem  for finding the optimum power allocation coefficients that minimize the ASER. Then,  we propose an effective sub optimum solution to the formulated non-convex optimization  problem   using SCP with trust region algorithm.
\item Extensive Monte Carlo simulations are carried out in order to validate the derived analytic expressions and to demonstrate the ASER performance under different spectral efficiency levels and LEDs separations.
\end{itemize}

The remainder of the paper is organized as follows: Section
II describes the channel and system model of the proposed APQ-SM based VLC
downlink network. Section III presents the proposed decoding process,  the derivation of the  analytic expressions for the corresponding ASER, and the formulation of the optimization problem. 
Numerical results and useful 
related discussions are presented in Section IV, whereas closing
remarks are provided in Section V.

\textit{Notations:} Throughout the paper, and unless otherwise stated, boldface uppercase and lowercase represent matrices and vectors, respectively, whereas $(\cdot )^{T}$ denotes the transpose operation. Moreover, $||\cdot||$ denotes the Euclidean norm and $|\cdot|$ represents the absolute value. 
Also, $\text{tr}(\mathbf{A})$ denotes the trace of a matrix $\mathbf{A}$, while diag$(\mathbf{x})$ is a diagonal matrix whose entries are the vector $\mathbf{x}$. Finally, $\mathbf{0}$  accounts for  the all-zeros vector.

\begin{figure*}[ht]
    \centering
    \includegraphics [width=16cm, height = 8cm] {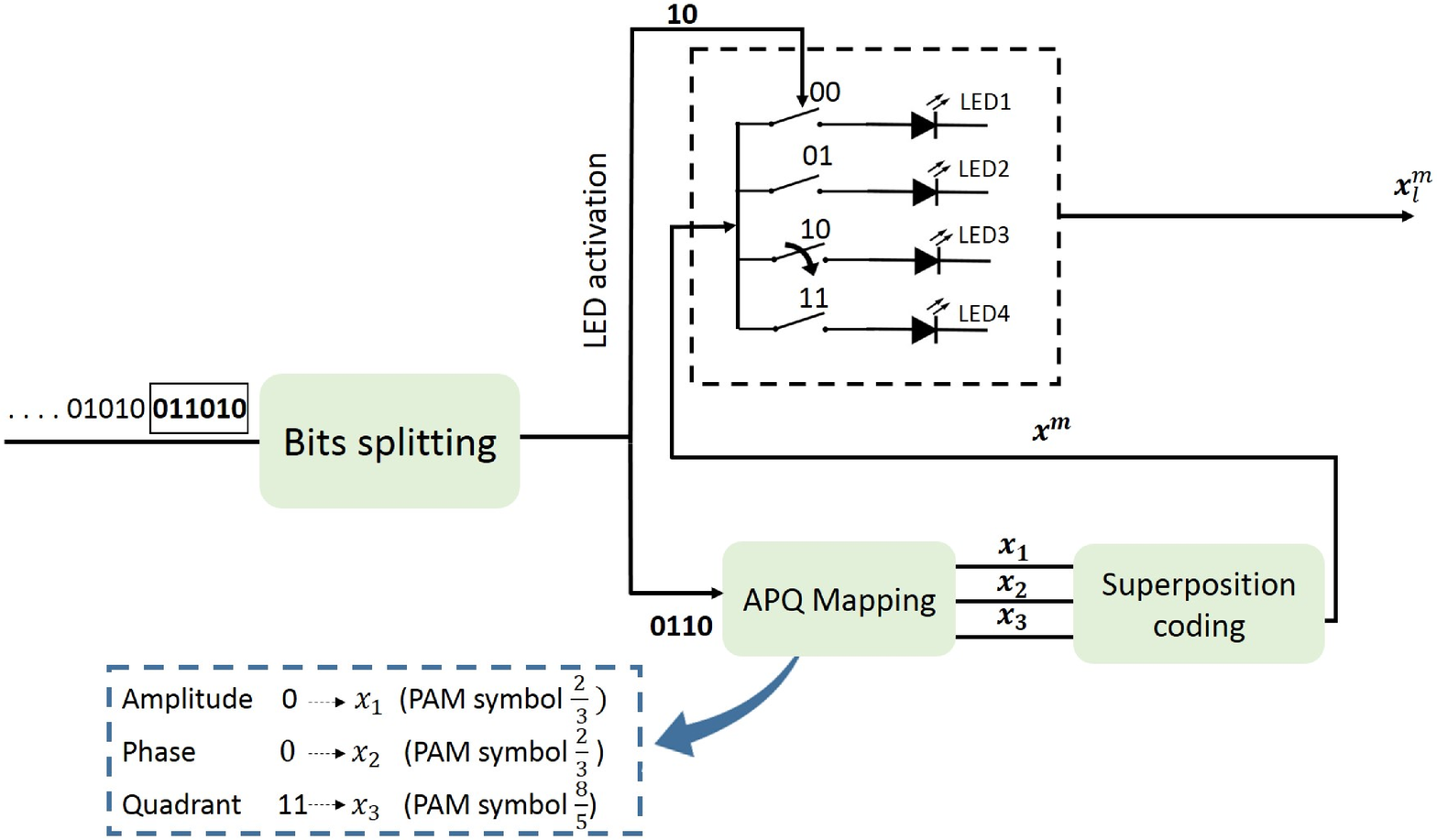}
    \caption{\textcolor{black}{APQ-SM transmitter.}}
    \label{model}
\end{figure*}

\section{System and Channel Models}

We consider an indoor VLC MIMO downlink system which consists of  $N_{t}$ transmit LEDs, assumed to be power of 2, and a user with $N_r$ PDs. 
Also, and without loss of generality, the dimensions of the room are considered to be  $3\;m\times3\;m\times3\;m$. 
To this effect, the transmitted bit stream is divided into two parts: i) the first part consists of  $\text{log} _2({N_t })$ bits, which are  used to select the  index of the active LED; ii) the second part consists of the other $\text{log}_2{(M)}$ bits, which are modulated using $M$-ary APQ modulation. The considered APQ-SM transmitter is illustrated in detail in Fig. \ref{model}.

Based on the above, in order to construct the $m^{\rm th}$ symbol in $M$-ary APQ modulation, the $\text{log}_2(M)$ bits are divided into three different parts, namely, the amplitude, phase, and  quadrant components. Then, each component is modulated using an $M_i$-ary PAM,
 as follows:
 \begin{equation}
     x^m_i \in \bigg\{\frac{2 }{M_i+1}, \frac{4}{M_i+1},\cdots,\frac{2\;M_i }{M_i+1}\bigg\},  \quad   i=1,2,3. 
     \label{symbols}
 \end{equation}
\textcolor{black}{It is worth  noting that the symbols in (\ref{symbols}) are chosen in order to maintain a fixed average optical power, and hence, fixed illumination in the room. }Subsequently, all components are superimposed in power domain and are transmitted simultaneously to fit with  the constraints imposed by the nature of VLC systems. Therefore, the $m^{th}$ APQ symbol ${x}^m$ can be represented as  \cite{ABUMARSHOUD2021101440}
\begin{equation}
    {x}^m= {p_1} x^m_1 +{p_2} x^m_2 +{p_3} x^m_3,
\end{equation}
where $p_1, p_2, \text{and } p_3$ represent the optical power allocation for each component. It is worth noting that we assumed that
 \begin{equation}
 \sum_{i=1}^3 p_i=P_{\rm opt},
 \end{equation}
 and $ p_i < p_{i-1}, \; i=2,3$, where $P_{\rm opt}$ denotes the mean optical power of the corresponding LED. Based on this, the average symbol energy, $E_s$, can be expressed as follows:
 \begin{equation}
 E_s= \gamma^2 P_{\rm opt}^2 \; T_s,
  \end{equation}
 where $\gamma$ represents the optical-to-electrical conversion factor, and $T_s$ denotes the symbol duration.
 
The concept of the proposed modulation scheme is demonstrated in more detail in  Fig. \ref{model}. This  represents  an APQ-SM system with an indicative spectral efficiency of $\eta = 6\; \text{bpcu}$, which is realized by having 4 LEDs at the transmitter side in order to transmit 2 bits in the spatial domain and 4 bits for the $16$-APQ symbol. Then, the $16$-APQ symbol is constructed as follows: the 4 bits are divided into three parts; the first bit is used to represent one of the two amplitudes ($\text{i.e., } L_1, L_2$) through 2-PAM modulation, the next two bits represent the four quadrant  through 4-PAM modulation, and  the last bit is used to represent the two phases ($\text{i.e., } \theta_1, \theta_2$) through 2-PAM modulation. Then, each resulted PAM symbol is allocated a fraction of the total power and finally they are superimposed and transmitted from the active LED.  \\
\indent Based on the above, the received information signal can be expressed as follows:
%
\begin{equation}
\mathbf{y}=  {\gamma}  \mathbf{H}\mathbf{x}_\ell^m+\mathbf{z},
\label{RX}
\end{equation}
where $\mathbf{x}_\ell^m$ is the $m^{\rm th}$ APQ symbol transmitted from the $\ell^{\rm th}$ LED, which is represented as  
 \begin{equation}
     \mathbf{x}_\ell^m = [0,\cdots,0, x^m,0,\cdots,0]. 
 \end{equation}
Moreover, $\mathbf{z}$ represents the additive white Gaussian noise (AWGN) vector, where each component has zero mean and variance $\sigma^2 = N_0  B$, with $N_0$ denoting the noise power spectral density and $B$  the channel bandwidth \cite{Ghassem2019}. Additionally, $\mathbf{H}_{N_r\times N_t}$ denotes the channel gain matrix between the LEDs and the PDs,  in which only the line-of-sight (LoS) components are considered.
Based on this, each component of $\mathbf{H}$ can be  expressed as {\cite{Komine2004}}
\begin{equation}
\label{eq:channel}
\small 
 h_{r\ell}= \left\{\begin{matrix}
\frac{A}{d^2_{r\ell}} R_{o}(\varphi_{r\ell}) T_{s}(\phi_{r\ell}) g(\phi_{r\ell}) \cos(\phi_{r\ell}),  & 0 \leqslant \phi_{r\ell} \leqslant \phi_{c}, \\
0, & \text{otherwise},
\end{matrix}\right.
\end{equation}
where $A$ denotes the PD area, $d_{r\ell}$ is the distance between the $\ell^{\rm th}$ LED and the $r^{\rm th}$ PD, $\varphi_{r\ell}$ is the angle of transmission from the $\ell^{\rm th}$ LED to the $r^{\rm th}$ PD, $\phi_{r\ell}$ is the incident angle with respect to the receiver, and $\phi_{c}$ is the field of view (FoV) of the PD. Furthermore, $T_{s}(\phi_{r\ell})$ and  $g(\phi_{r\ell} )$ denote the gains of the optical filter and concentrator, respectively, whereas $g(\phi_{r\ell} )$ can be expressed as follows:
\begin{equation}
\label{eq:g}
g(\phi_{r\ell})= \left\{\begin{matrix}
\frac{n^2}{\sin^2({\phi_{c}})},  & 0 \leqslant {\phi_{r\ell}} \leqslant \phi_{c},\\
0, &  {\phi_{r\ell}}>\phi_{c}.
\end{matrix}\right.  
\end{equation}
In the above,  $n$ represents the corresponding refractive index, and $R_{o}(\varphi_{r\ell})$ denotes the Lambertian radiant intensity, namely 
\begin{equation}
\label{eq:R}
R_{o}(\varphi_{r\ell})= \frac{\nu+1}{2\pi} \left(\cos(\varphi_{r\ell})\right)^{\nu},
\end{equation}
with $\nu$ representing  the order of the Lambertian emission, which is  expressed as
\begin{equation}
\label{eq:m}
\nu = \frac{-\ln{(2)}}{\ln\left({\cos(\varphi_{1/2})}\right)}, 
\end{equation}
given that $\varphi_{1/2}$ is the LED semi-angle at half power \cite{Mesleh2011_2}. 

\section{Proposed ML Decoding }

Following from the above, this section is devoted in the  quantification of the  performance of the proposed system model by means of analyzing the achievable symbol error rate performance.
To this end, assuming maximum likelihood (ML) detection is utilized at the receiver side, the  detector performs joint detection to deduce the received signals over the space and signal domains. In particular, the receiver detects jointly the LED index and the transmitted  APQ symbol according to the following criterion: 
%
\begin{equation}
\mathbf{x}^{\hat{m}}_{\hat{\ell}}=\arg\min_{\forall{\mathbf{x}^{\tilde{m}}_{\tilde{\ell}}}\in  \chi}\left\Vert\mathbf{y}- \gamma\mathbf{H}{\mathbf{x}^{\tilde{m}}_{\tilde{\ell}}}\right\Vert^2,
\end{equation}
where  $\mathbf{x}^{\hat{m}}_{\hat{\ell}}$ is the estimated APQ-SM symbol that minimizes the distance between the received signal and all potential APQ-SM symbols. 
It is recalled here that joint ML detection requires the search over all ${M \times  {N_t} }$ combinations of the indices and symbols, which results in an increased receiver complexity. Therefore, the upper bound on the achievable SER  using joint detection, $P_{e, \rm joint}$, is expressed as 
\begin{equation}
    \centering
 P_{e, \rm joint} \leqslant  \frac{1}{N_t \; M}{\sum\limits_{\substack{ \forall (m,\ell) }}\sum \limits_{\substack{ \forall (\hat{m}, \hat{\ell}) \\ \neq\\ ({m},{\ell})}} } Q \left(\sqrt{ \frac{\gamma^2}{4 \sigma^2} \norm{\mathbf{H}(\mathbf{x}^m_\ell-\mathbf{x}_{\hat{\ell}}^{\hat{m}}) }^2} \right), 
 \label{JOINT}
\end{equation}
where $Q(\cdot)$ denotes the one dimensional Gaussian $Q-$function \cite{prudnikov1986integrals}. 
As already mentioned, it is of paramount importance to achieve a reduced complexity receiver in the   considered VLC system. 
To that end,  we propose a new detection mechanism in which the detection process is performed over two stages. \textcolor{black}{Hence, this yields a reduced complexity in terms of number of iterations, i.e., the search space is reduced to  $M+ \lfloor{N_t \choose 1}\rfloor_{2^p}$}. The first stage comprises a conditional ML detection, in which, conditioned on $\mathbf{x}^{{m}}_{\tilde{\ell}}$, the receiver detects only the index of the transmitting LED, namely
%
%
\begin{equation}
\hat{\ell}=\arg\min_{\forall\tilde{\ell}}\bigg(\min_{\forall m} \left\Vert \big[ \big(\mathbf{y}- {\gamma}\mathbf{H} \mathbf{x}^{{m}}_{\tilde{\ell}} \big) \big|\mathbf{x}^{{m}}_{\tilde{\ell}} \big ]\right\Vert^2 \bigg).
\label{index}
\end{equation}
Then, in the second stage the receiver uses the detected index in order to detect the transmitted symbols. 
To this effect and assuming perfect detection of the $\ell^{\rm th}$ LED index, the estimated symbol can be evaluated as follows:

\begin{equation}
\mathbf{x}^{\hat{m}}_\ell=\arg\min_{\forall \tilde{m}} \bigg\Vert {\mathbf{y}- \gamma \mathbf{H} \; \mathbf{x}^{\tilde{m}}_\ell } \bigg \Vert^2.
\label{symbol}
\end{equation}
Capitalizing on the above, the next section is devoted in the evaluation of the achievable  ASER performance of the proposed receiver.

\subsection{Average SER Analysis}
The probability of incorrect detection given that the $\ell^{\rm th}$ LED  is selected can be determined by 
\begin{equation}
P^{\ell}_e =  P_x(\mathbf{x}_\ell^m|\ell\neq\hat{\ell}). P_{\ell} + P_x(\mathbf{x}_\ell^m|\ell=\hat{\ell}) . (1-P_{\ell}),
\label{total_error}
\end{equation}
where $P_{\ell}$ is probability of incorrect index detection, whereas  $P_x(\mathbf{x}_\ell^m|\ell=\hat{\ell})$ and $P_x(\mathbf{x}_\ell^m|\ell\neq\hat{\ell})$ denote the probability of incorrect symbol detection conditioned on correct and incorrect index detection, respectively. Due to mathematical intractability of (\ref{total_error}) and given that $ P_x(\mathbf{x}_\ell^m|\ell\neq\hat{\ell})$ is rather large, close to unity, the  error probability for the $\ell^{\rm th}$ combination can be accurately simplified to the following 
\begin{equation}
P^\ell_e \leqslant  P_{\ell} + P_x(\mathbf{x}_\ell^m|\ell=\hat{\ell}) . (1-P_{\ell}).
\label{approx}
\end{equation}
\noindent
Therefore, it is evident that the ASER can be obtained from (\ref{approx}) subject to analytic derivation of two explicit  expressions, for  $P_{\ell}$ and $ P_x(\mathbf{x}_\ell^m|\ell=\hat{\ell}) $ and subsequent averaging over all possible LEDs activation.
Based on this, the following closed form expression for the ASER for $P_{\ell}$ is obtained  
\begin{equation}
\centering
P_\ell =  \frac{1}{M}\sum\limits_{\forall m}  {Q} \left(\sqrt{ \frac{\gamma^2}{4 \sigma^2} D^2(\mathbf{x}^m_{\ell})} \right),
\label{union_index}
\end{equation}
\noindent
where $ D(\mathbf{x}_{\ell}^m)$ is the minimum distance between the symbol $\mathbf{x}^m_{\ell}$ transmitted from the $\ell^{\rm th}$ LED index and all possible symbols that can be transmitted from all other LEDs, namely 
%
\begin{equation}
\centering
   D(\mathbf{x}_{\ell}^m)= \min_{\substack{\forall \hat{\ell}, \ell \neq \hat{\ell}\\ \forall \hat{m}} } \norm{\gamma\mathbf{H}( \mathbf{x}_\ell^m - \mathbf{x}^{\hat{m}}_{\hat{\ell}} )}. 
\end{equation}
\noindent
Based on this and utilizing the union bound on the SER, we evaluate the probability of incorrectly detecting the symbol, $P_x(\mathbf{x}_\ell^m|\ell=\hat{\ell})$, as follows:
\begin{equation}
\centering
P_x(\mathbf{x}_\ell^m|\ell=\hat{\ell}) =  \frac{1}{ M}\sum\limits_{\forall m} \sum\limits_{\substack{\forall \hat{m}\\ m \neq \hat{m}}} P(\mathbf{x}^m_\ell \longrightarrow {\mathbf{x}_\ell^{\hat{m}}}),
\label{union_symbol_unsimplified}
\end{equation}
where $P(\mathbf{x}^m_\ell \longrightarrow {\mathbf{x}_\ell^{\hat{m}}})$ denotes the pairwise error probability (PEP) of decoding  $\mathbf{x}^m_\ell$ as $\mathbf{x}_\ell^{\hat{m}}$,  which is represented as 
\begin{equation}
P(\mathbf{x}^m_\ell \longrightarrow {\mathbf{x}_\ell^{\hat{m}}})  =  Q \left(\sqrt{ \frac{\gamma^2}{4 \sigma^2} \norm{\mathbf{H}(\mathbf{x}^m_\ell-\mathbf{x}_\ell^{\hat{m}}) }^2} \right). 
 \label{PEP_x}
\end{equation}
Therefore, by substituting (\ref{PEP_x}) in (\ref{union_symbol_unsimplified}), the  ASER for incorrectly detecting the symbol $P_x(\mathbf{x}_\ell^m|\ell=\hat{\ell})$ is given by the following  analytic expression
\begin{dmath}
\centering
P_x(\mathbf{x}_\ell^m|\ell=\hat{\ell}) =  \frac{1}{ M}\sum\limits_{\forall m} \sum\limits_{\substack{\forall \hat{m}\\ m \neq \hat{m}}}  Q \left(\sqrt{ \frac{\gamma^2}{4 \sigma^2} \norm{\mathbf{H}(\mathbf{x}^m_\ell-\mathbf{x}_\ell^{\hat{m}}) }^2} \right).
\label{union_symbol}
\end{dmath}

\noindent
Finally, using (\ref{approx}), (\ref{union_index}) and (\ref{union_symbol}) and averaging over all possible LEDs activation, the total upper bound on the corresponding SER can be expressed according to  (\ref{final_expression}). 

To the best of the authors' knowledge, the offered analytic results have not been previously reported in the open technical  literature.  In what follows, these results are employed in the determination of effective power allocation strategies in the considered APQ-SM based VLC system.  
\begin{figure*} 
\centering
\begin{equation}
\begin{split}
\label{final_expression} 
 P_e \leqslant    \frac{1}{N_t} \sum_{\forall \ell} & \left[    \frac{1}{M}\sum\limits_{\forall m}  {Q} \left(\sqrt{ \frac{\gamma^2}{4 \sigma^2} D^2(\mathbf{x}^m_{\ell})} \right) + \frac{1}{ M}\sum\limits_{\forall m} \sum\limits_{\substack{\forall \hat{m}\\ m \neq \hat{m}}}  Q \left(\sqrt{ \frac{\gamma^2}{4 \sigma^2} \norm{\mathbf{H}(\mathbf{x}^m_\ell-\mathbf{x}_\ell^{\hat{m}}) }^2} \right) \right. \\
 & \left. -  \frac{1}{M}\sum\limits_{\forall m}  {Q} \left(\sqrt{ \frac{\gamma^2}{4 \sigma^2} D^2(\mathbf{x}^m_{\ell})} \right) \times \frac{1}{ M}\sum\limits_{\forall m} \sum\limits_{\substack{\forall \hat{m}\\ m \neq \hat{m}}}  Q \left(\sqrt{ \frac{\gamma^2}{4 \sigma^2} \norm{\mathbf{H}(\mathbf{x}^m_\ell-\mathbf{x}_\ell^{\hat{m}}) }^2} \right) \right].
\end{split}
\end{equation}
\hrulefill
\end{figure*}


\subsection{Power Allocation Optimization}

It is recalled that the performance of the proposed scheme is largely dependent upon the factors relating to the corresponding power allocation strategy. Based on this, in this section we formulate an optimization problem that aims to minimize the ASER obtained from  the performed joint detection. 
To that end, we first expand the norm to represent $P_{e,joint}$ in (\ref{JOINT}) in terms of the $p_i$ parameters   in (\ref{eq19}), 
\begin{figure*}[t]
\begin{equation}
    \begin{split}
 P_{e, \rm joint} \leqslant  \frac{1}{N_t \; M}\sum\limits_{\forall \ell}\sum\limits_{\forall m} \sum \limits_{\substack{\forall \hat{\ell}}} \sum\limits_{\forall \hat{m}} Q{ \left({ \frac{\gamma}{2 \sigma} }\sqrt{ \sum_{r=1}^{N_r} \Bigg | \bigg ( [p_1 x_1^{\hat{m}}+p_2 x_2^{\hat{m}} +p_3 x_3^{\hat{m}}] h_{r\hat{\ell}} \bigg ) -\bigg( [p_1 x_1^{{m}}+p_2 x_2^{{m}} +p_3 x_3^{{m}}] h_{r{\ell}}   \bigg) \Bigg |^2  } \right), }\\ (m,\ell) \neq (\hat{m},\hat{\ell}).
     \end{split}
 \label{eq19}
\end{equation}
\hrulefill
\end{figure*}
which can be further simplified as to  (\ref{eq20}), at the top of the next page,
\begin{figure*}[ht]
\centering
\begin{equation}
P_{e, \rm joint} \leqslant  \frac{1}{N_t \; M}\sum\limits_{\forall \ell}\sum\limits_{\forall m} \sum \limits_{\substack{\forall \hat{\ell}}} \sum\limits_{\forall \hat{m}} Q{ \left({ \frac{\gamma}{2 \sigma} } \sqrt{\sum_{r=1}^{N_r} \bigg | p_1 \Delta^r_{1}(m,\ell,\hat{m},\hat{\ell})+p_2 \Delta^r_{2}(m,\ell,\hat{m},\hat{\ell}) +p_3 \Delta^r_{3}(m,\ell,\hat{m},\hat{\ell}) \bigg |^2   }\right), } \;(m,\ell) \neq (\hat{m},\hat{\ell}).
\label{eq20}
\end{equation}
\hrulefill
\end{figure*}
where 
\begin{align}
\Delta^r_i(m,\ell,\hat{m},\hat{\ell}) =  x_i^{\hat{m}} h_{r\hat{\ell}}-x_i^{{m}} h_{r{\ell}}, 
\end{align}
with $\; i=1,2,3$ and  $r=1,\cdots,N_r$. 
To this effect and assuming a predetermined spectral efficiency, we formulate the following minimization problem:
\begin{subequations}
\small
	\begin{align}
	\min_{\mathbf{p}}    \quad & 
	{\frac{1}{N_t \; M}   \sum\limits_{\forall \ell}\sum\limits_{\forall m} \sum \limits_{\substack{\forall \hat{\ell} }} \sum\limits_{\forall \hat{m}} Q{ \left({ \frac{\gamma}{2 \sigma} } \sqrt{ \sum_{r=1}^{N_r} \big |\mathbf{p}^T \mathbf{\Delta}^r(m,\ell,\hat{m},\hat{\ell}) \big |^2  } \right) } } \tag{P1} \\
	\label{c1}
	\text{s.t.}\quad & {\text{tr}(\text{diag}(\mathbf{p}))= P_{opt},} \tag{P1.a}\\
	\label{c2_1} & {\mathbf{p}\geq \mathbf{0},}\tag{P1.b}\\
	\label{c2_1} &  p_i\leqslant p_{(i-1)}, \; \; i=2,3,\tag{P1.c}
     \end{align}
   \end{subequations}
\noindent
where  
$ \scriptstyle \mathbf{\Delta}^r(m,\ell,\hat{m},\hat{\ell}) = [\Delta^r_1(m,\ell,\hat{m},\hat{\ell}),\Delta^r_2(m,\ell,\hat{m},\hat{\ell}), \Delta^r_3(m,\ell,\hat{m},\hat{\ell})]$, whilst  $\mathbf{p}=[p_1,p_2,p_3]^T $ denotes the power allocation vector. 
It is noted here that the optimization problem in (P1) is non-convex due to the non-convexity of the objective function in $\mathbf{p}$. 
In what follows, we propose an effective sub-optimum solution to this problem with the aid of SCP algorithm, due to its simplicity and high efficiency \cite{mao2018,SAMPATH2018}. 
To that end, we re-describe the algorithm by first noting that  SCP solves a non-convex problem by repeatedly constructing a convex approximation to the problem around the current iterate $\mathbf{p}^{(l)}$. 
The sub-problem is then used to generate the new value of $\mathbf{p}^{(l+1)}$ that will be used in the next iteration. 
Based on this, there are different choices for the approximation function, namely:  first-order Taylor series, second-order Taylor series with positive semi-definite Hessian, inner convex approximation, or any other convex function that locally approximates the non-convex objective function at $\mathbf{p}^{(l)}$. In the present analysis, the objective function is replaced by its first order Taylor series expansion at each iteration $l$ around the point  $\mathbf{p}^{(l)}$. However, the convergence and the accuracy of the approximation are particularly sensitive to the $||\mathbf{p}-\mathbf{p}^{(l)}||$ term. Hence,  we incorporate the trust region method  to control the step size which is reliable and robust, whilst it also exhibits adequate convergence properties. 
Based on this, it follows that (P1) can be approximated at iteration $l$ by the following linear sub-problem: \par
\begin{subequations}
\small
	\begin{align}
	\min_{\mathbf{p} } & \quad 
	{ E^{(l)}} \tag{P2} \\
	\label{c1}
	\text{s.t.} \quad &
{\text{tr}(\text{diag}(\mathbf{ p}))= P_{opt},}\tag{P2.a}  \\
	\label{c2_1} & {\mathbf{ p}\geq \mathbf{0},} \tag{P2.b}\\
		\label{c2_1} &  p_i\leqslant p_{(i-1)}, \; \; i=2,3,\tag{P2.c}\\
	\label{c2_1} & ||\mathbf{p}-\mathbf{p}^{(l)} || \leqslant \delta^{(l)}, \tag{P2.d}
	\end{align}
\end{subequations}
where the constraint in (P2.d) is used to define the trust region and $\delta^{(l)}$ is the iteration-based trust region radius \cite{sheng2016,mao2016}. Additionally,  $E^{(l)}$ is the first-order Taylor approximation at the $l^{\rm th}$ iteration, which is expressed as
%
\begin{equation}
    E^{(l)} = (\mathbf{A}^{(l)})^T (\mathbf{p}-\mathbf{p}^{(l)})+ B^{(l)},
\end{equation}
where  $\mathbf{A}^{(l)}$ and $\mathbf{B}^{(l)}$ are given by the explicit expressions in equations  \eqref{term1} and \eqref{term2}. 
\begin{figure*} 
\centering
\begin{equation} \label{term1}
\mathbf{A}^{(l)}= \frac{-\gamma}{\sigma N_t \; M\; \sqrt{8\pi}}\sum\limits_{\forall \ell}\sum\limits_{\forall m} \sum \limits_{\substack{\forall \hat{\ell} }} \sum\limits_{\forall \hat{m}} e^{- \frac{1}{2} \bigg({ \frac{\gamma^2}{4 \sigma^2} } \sum_{r=1}^{N_r} \big |(\mathbf{p}^{(l)})^T \mathbf{\Delta}^r(m,\ell,\hat{m},\hat{\ell}) \big |^2   \bigg)}    \frac{\sum_{r=1}^{N_r}  \bigg( (\mathbf{p}^{(l)})^T \mathbf{\Delta}^r(m,\ell,\hat{m},\hat{\ell})\bigg) \mathbf{\Delta}^r(m,\ell,\hat{m},\hat{\ell})}{\sqrt{\sum_{r=1}^{N_r} \bigg |(\mathbf{p}^{(l)})^T \mathbf{\Delta}^r(m,\ell,\hat{m},\hat{\ell}) \bigg |^2  }}.
\end{equation}
\hrulefill
\end{figure*}
\begin{figure*} 
\centering
\begin{equation} \label{term2}
    B^{(l)}= \frac{1}{N_t \; M}\sum\limits_{\forall \ell}\sum\limits_{\forall m} \sum \limits_{\substack{\forall \hat{\ell} }} \sum\limits_{\forall \hat{m}} Q{ \left({ \frac{\gamma}{2 \sigma} } \sqrt{ \sum_{r=1}^{N_r} \bigg |(\mathbf{p}^{(l)})^T \mathbf{\Delta}^r(m,\ell,\hat{m},\hat{\ell}) \bigg|^2   }\; \right).}  
\end{equation}
\hrulefill
\end{figure*}
It is recalled here that  in trust region method, the new iteration is generated within a close region around the current point. Then, the radius of this region is adjusted according to an indicator function that determines how accurately the approximate model fits  the original problem. 
Based on this, in order to decide on the value of $\delta^{(l+1)}$, we utilize the following merit function for the actual problem
\begin{dmath}
    f_a(\mathbf{p}) =	{\frac{1}{N_t \; M}\sum\limits_{\forall \ell}\sum\limits_{\forall m} \sum \limits_{\substack{\forall \hat{\ell} }} \sum\limits_{\forall \hat{m}} Q{ \left({ \frac{\gamma}{2 \sigma} } \sqrt{ \sum_{r=1}^{N_r} \big |\mathbf{p}^T \mathbf{\Delta}^r(m,\ell,\hat{m},\hat{\ell}) \big |^2  } \right),} } 
\end{dmath}
along with the following merit function for the approximated sub-problem 
\begin{dmath}
    f_p(\mathbf{p})= E^{(l)}.
\end{dmath}
To this effect and  in order to decide on the value of $\delta^{(l+1)}$, the following relative   performance metric is utilized 
\begin{equation}
    r^{(l)}=\frac{f_a(\mathbf{p}^l)-f_a(\mathbf{p}^{l+1})}{f_p(\mathbf{p}^l)-f_p(\mathbf{p}^{l+1})}.
\end{equation}
Therefore, based on the value of $r^{(l)}$ the radius of the trust region is either, increased, decreased, or maintained unchanged. Additionally,  the value of $r^{(l)}$ assists in deciding on whether to accept or reject the estimated point as a reference in the next iteration. Importantly, a common approach on this is to define three real numbers, $\alpha_0< \alpha_1< \alpha_2 \in (0,1)$, that split the range of the possible values of $r^{(l)}$ into four segments. Then, the value of the trust region radius and the reference point for the next iteration are updated according to the rule in the pseudo-code for the SCP detailed in Algorithm 1, where the parameters $\alpha, \beta >0$ are user selected values and $\epsilon$ denotes the termination tolerance.  

\textcolor{black}{
In Algorithm 1, the complexity of the power allocation is mainly associated with the SCP algorithm which is determined by the number of iterations that are required for convergence as well as the complexity for each iteration. Note that the number of iterations required by SCP is $\mathcal{O}\big(\sqrt{N_1}\; \text{log}_2(1/\epsilon )\big)$, where $\epsilon$ is the accuracy of SCP and $N_1$ is the number of constraints which is $2K-1$; here $K$ denotes the number of variables in (P2). At each iteration, (P2) is solved with complexity of  $\mathcal{O}(N_1\; K^2)$ \cite{LOBO1998}. Therefore, the overall complexity associated with the SCP is $\mathcal{O}( K^{1.5} \; \text{log}_2(1/\epsilon))$, which means that the sub-optimal solution is computed in a polynomial time \cite{9258414,Boyd2004}.} 

%
\begin{algorithm}[ht]
 \text{\textbf{Initialization}:}{{ set  $\mathbf{p}^{(0)}$,$\delta^{(0)}$, $\alpha_0,\alpha_1,\alpha_2, \alpha, \beta,$ and tolerance $\epsilon$ }} \\
  \For{l = 0 \text{until meeting termination condition}}{
 Obtain   $\mathbf{p}^{(l+1)}$ by solving (P2) using $\mathbf{p}^{(l)}$:\\
  \eIf{$| \mathbf{p}^{(l+1)}-\mathbf{p}^{(l)} |\leq \epsilon$  , or$\;N_{max}$ is reached}{
  Break\;
   }{
   \If{$r^{(l)}\geq \alpha_2$}{
$\delta^{(l+1)}=\beta \delta^{(l)}$, and accept  $\mathbf{p}^{(l+1)}$
   }
      \If{$\alpha_1 \leqslant r^{(l)} < \alpha_2$}{
$\delta^{(l+1)}= \delta^{(l)}$, and accept  $\mathbf{p}^{(l+1)}$
   }
    \If{$\alpha_0 \leqslant r^{(l)} < \alpha_1$}{
$\delta^{(l+1)}= \delta^{(l)}/\alpha$, and accept  $\mathbf{p}^{(l+1)}$
   }
  \If{$r^{(l)} < \alpha_0$}{
$\delta^{(l+1)}= \delta^{(l)}/\alpha$, and reject  $\mathbf{p}^{(l+1)}$, $\mathbf{p}^{(l)}$ is retained
   } 
 }{    Set $l = l + 1$} }
 \caption{ SCP algorithm with trust region approach }
 \end{algorithm}
\indent 
It is noted here that evaluating the SER performance with respect to the average  received electrical signal-to-noise ratio (SNR) would disregard the individual path loss of different setups and the particular activated LED. Therefore, in order to provide a more fair comparison, we evaluate the SER performance of the proposed scheme with respect to the transmit SNR, which is defined as the ratio of the average symbol energy against the noise power
spectral density  $ E_s/N_0$ \cite{Fath2013,Fath_20132}. 
It is also emphasized that the value of $ E_s/N_0$ in VLC systems is considerably greater  than that of  counterpart RF systems \cite{He2015,He2015_2}. This difference is attributed to the small value associated with the noise power spectral density $N_0$ \cite{Yin2016}. 
More specifically, by neglecting the photodetector dark current, the $N_0$ term is expressed as $N_0 \simeq q\;I_B $, where $q=1.6e{-19}$ is the charge of electron and $I_B$ denotes the involved background noise current that typically takes values in the order of $\mu A$ \cite{Stavridis_2015,Ghassem2019} and the references therein.   

 \section{Numerical Results and Insights}
 
 In what follows, we exploit the   offered results in the previous sections to quantify the achievable  ASER of the proposed modulation scheme in the considered VLC  setup. 
 The presented analytic representations are thoroughly corroborated by extensive results from respective Monte  Carlo  simulations, which justify the validity of the analytic derivations and the proposed SCP based suboptimum algorithm for the corresponding  power allocation. 
The obtained analytic and simulation  results are subsequently compared with respective results corresponding to the conventional SM with PAM scheme as well as to the MA-SM scheme.
 Without loss of generality, these analyses and comparisons are carried out  for the case of an indoor environment with dimensions  $3\; m\times 3\; m\times3\; m$, which includes  $N_t=4$ high brightness white LEDs attached to the ceiling at a height of $2.5\; m$ and with a $d_{\rm TX}$  separation distance between them. 
 Additionally, we assume that the receiver has $N_r=4$ PDs placed at a height of $0.75\; m$. 
 All   parameters used in the simulation as well as the LEDs and PDs locations are summarized in Tables \ref{TableI} and \ref{TableII}, respectively.
 
\begin{table}[ht]
\centering
\caption{Simulation parameters.}
\begin{tabular}[c] {p{3cm} p{1.5cm} p{2.5cm}}
\hline 
\hline
{Parameter}&{Symbol}&{Value} \\
\hline 
Room dimensions&- &$3\;m\times3\;m\times3\;m$ \\ 
LED beam angle& $\varphi_{1/2}$ &	$15^o$\\ 
PD area& $A$ &	$1\; cm^2$\\ 
Gain of optical filter&	$T_s(\phi_{r\ell})$  &1\\ 
FoV of PD & $\phi_c$	& $15^o$\\
Optical-to-electrical conversion factor& $\gamma$ &	$1\; A/W$\\ 
LED optical power& $P_{opt}$ &	$1 \;W$\\ \hline  \hline 
\end{tabular}
\label{TableI}
\end{table}

\begin{table}[H]
\centering
\caption{LEDs and PDs locations.}
\begin{tabular} [c]{p{3cm} p{3cm}}
\hline \hline
{LEDs locations}&{PDs locations}\\($d_{TX}=0.2\;m$) &\\  \hline 
LED  1: [1.6 1.6 2.5]& PD 1: [1.55 1.55 0.75]\\ 
LED  2: [1.4 1.6 2.5]&PD 2: [1.45 1.55 0.75]\\ 
LED  3: [1.6 1.4 2.5]&PD 3: [1.55 1.45 0.75]\\ 
LED  4: [1.4 1.4 2.5]&PD 4: [1.45 1.45 0.75]\\ \hline \hline 
\end{tabular}
\label{TableII}
\end{table}

 \begin{figure}[ht]
\centering
\includegraphics[width=250pt]{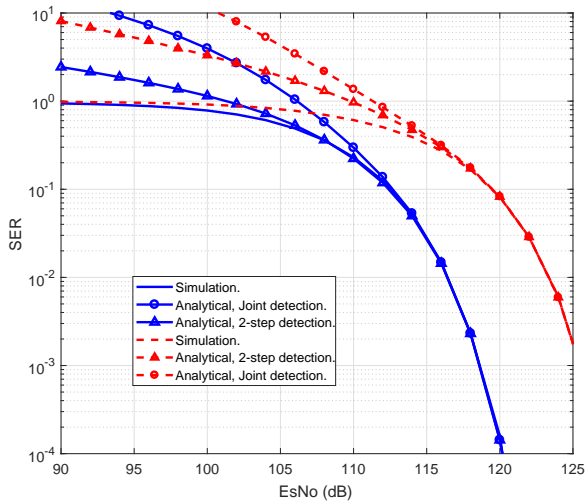}
\caption{ Comparison between the two-step and joint detection methods. Solid lines correspond to $\eta = 6 \rm \, bpcu$, whereas dashed lines correspond to $\eta = 8\rm \, bpcu$. }
\label{Fig2}
\end{figure}
 
 \begin{figure}[ht]
\centering
\includegraphics[width=250pt]{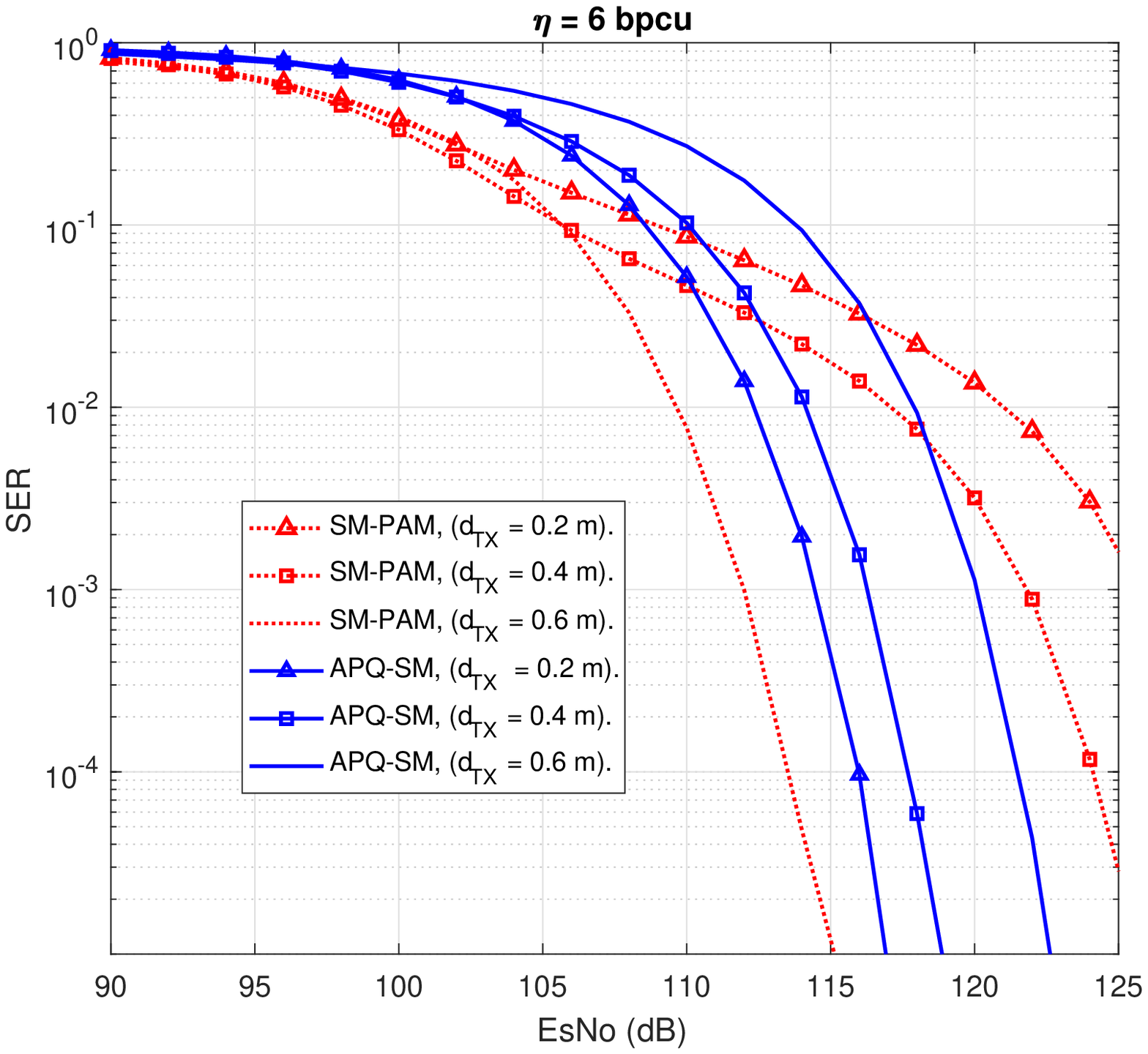}
\caption{ Comparison between APQ-SM and SM-PAM \textcolor{black}{in \cite{Fath2013}} for spectral efficiency 6 bpcu. }
\label{Fig3}
\end{figure}

 \begin{figure}[ht]
\centering
\includegraphics[width=250pt]{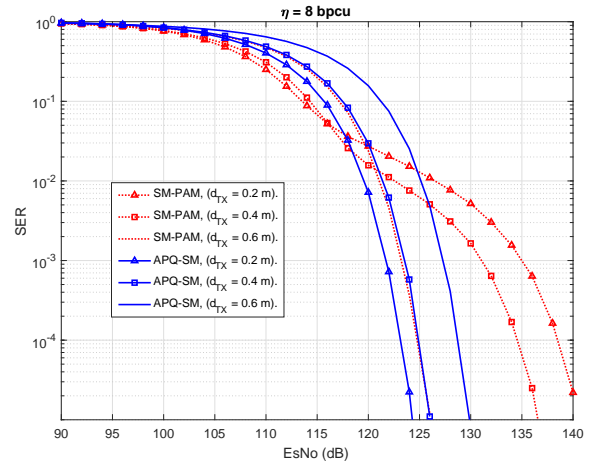}
\caption{ Comparison between APQ-SM and SM-PAM \textcolor{black}{in \cite{Fath2013}} for spectral efficiency 8 bpcu. }
\label{Fig4}
\end{figure}
Based on the above, we first validate the derived analytic expressions  for the upper bound on the  ASER in Section III. Fig. \ref{Fig2} illustrates the analytical and Monte Carlo results for the SER  with regard to the transmit SNR for two different spectral efficiency values, namely $\eta= 6  \rm \, bpcu$ and $\eta = 8   \rm \, bpcu$. It is clearly observed that the proposed upper bounds  are particularly tight from moderate SNR values to the  high SNR regime with respect to the corresponding  Monte Carlo simulation. 
It is also noticed that since the channel gains are in the order of $10^{-4}$, the received SNR will experience an about $80 \rm dB$ shift with respect to the transmit SNR, which corresponds to the typical values in VLC systems. In addition, it can be noticed that the two-step detection process is tighter at the low SNR regime, which is in fact because the detection is performed in  two stages.
 \begin{figure}[ht]
\centering
\includegraphics[width=250pt]{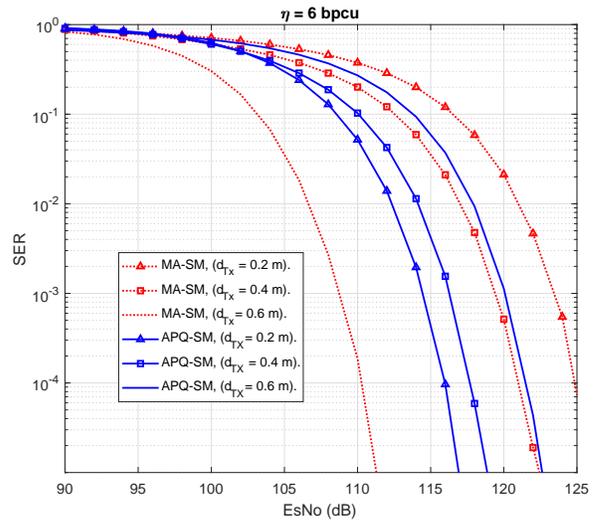}
\caption{ Comparison between APQ-SM and MA-SM \textcolor{black}{in \cite{7416970}} for spectral efficiency 6 bpcu. }
\label{Fig5}
\end{figure}
 
  \begin{figure}[ht]
\centering
\includegraphics[width=250pt]{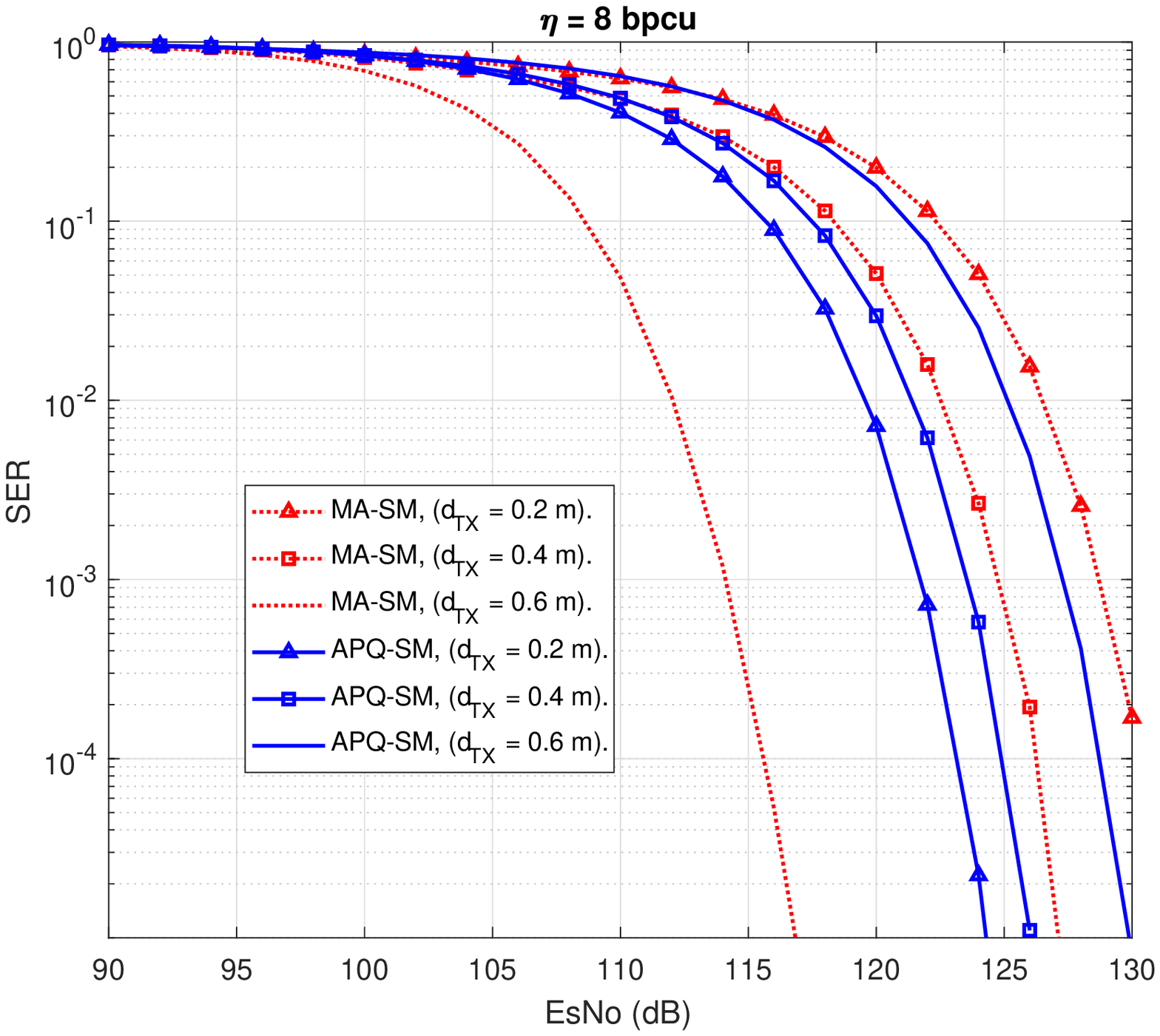}
\caption{ Comparison between APQ-SM and MA-SM \textcolor{black}{in \cite{7416970}} for spectral efficiency 8 bpcu. }
\label{Fig6}
\end{figure}
Next, we examine the SER performance
of the proposed modulation scheme and compare it with the SM-PAM scheme. 
Fig. \ref{Fig3} and Fig. \ref{Fig4} illustrate the achievable SER for indicative  spectral efficiencies of $\eta = 6 \rm \, bpcu$ and $\eta= 8\rm \, bpcu$, respectively. In addition, the effect of varying the $d_{\rm TX}$ distance between the transmitters on the x-axis and  y-axis  is quantified.
This reveals that the proposed APQ-SM modulation scheme appears to be more robust in terms of the separation between the transmitting LEDs for both values of spectral efficiency.  Furthermore, as the separation between the LEDs decreases, the achieved SER for the proposed modulation scheme is improved compared to the SM with PAM counterpart.
Therefore, it is evident that APQ-SM exhibits an overall enhancement on the achievable SER performance, particularly for higher spectral efficiency values. 
This can be justified by the fact that for a fixed number of LEDs in the considered room,  SM-PAM requires higher order PAM symbols to achieve the same spectral efficiency  that is achieved by the proposed APQ-SM scheme.

\begin{figure}[ht]
\centering
\includegraphics[width=250pt]{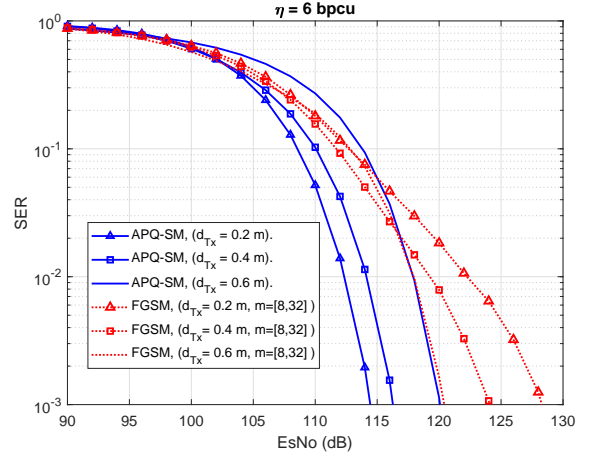}
\caption{\textcolor{black}{ Comparison between the proposed scheme and FGSM in \cite{9309358} for spectral efficiency 6 bpcu. }}
\label{Fig:FGSM_SE6}
\end{figure}

\begin{figure}[ht]
\centering
\includegraphics[width=250pt]{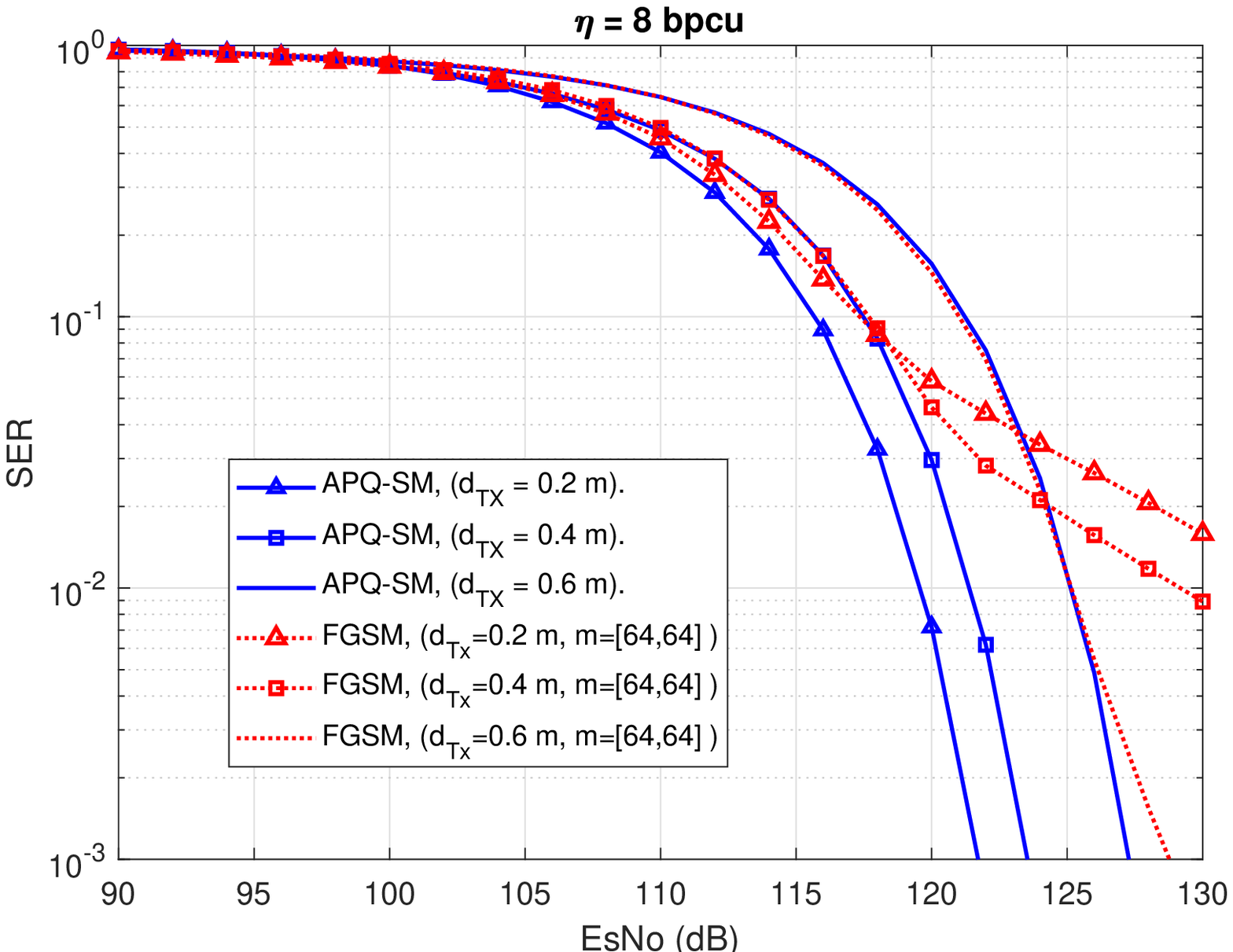}
\caption{\textcolor{black}{ Comparison between the proposed scheme and FGSM in \cite{9309358} for spectral efficiency 8 bpcu.} }
\label{Fig:FGSM_SE8}
\end{figure}

In the same context, Fig. \ref{Fig5} and Fig. \ref{Fig6} illustrate the SER comparison between the proposed APQ-SM scheme and the MA-SM counterpart for indicative  spectral efficiencies of  $\eta=6 \rm \, bpcu$ and $\eta=8\rm \, bpcu$, respectively. 
To that end, it is observed that since the LEDs combinations are used to transmit more information,   MA-SM exhibits better performance compared to the SM-PAM, where a single LED is activated. Nevertheless, it is also evident that the proposed APQ-SM scheme exhibits better performance for small LEDs separation, for the case of both considered spectral efficiencies. 
Therefore, it is overall concluded that the proposed APQ-SM scheme is more robust to the  separation distance of the LEDs.\\
\indent \textcolor{black}{In order to demonstrate the effectiveness of the proposed scheme, we have provided a  comparison with the recent work on flexible generalized SM (FGSM) in \cite{9309358}. For fairness, we have considered two fixed spectral efficiency values, i.e., 6 bpcu and 8 bpcu, as depicted in Figs. \ref{Fig:FGSM_SE6}-\ref{Fig:FGSM_SE8}, respectively. Furthermore, we assumed that the two schemes work under the same  $4\times4$ MIMO system configuration. For the FGSM, the spectral efficiency of 6 bpcu is achieved by considering two groups of LEDs, where the PAM constellation for the groups are chosen as $m=[8,32]$ and activating two LEDs to transmit the same PAM symbol. On the other hand, the spectral efficiency of 8 bpcu is achieved by choosing the PAM constellation for the groups as $m=[64,64]$ and activating two LEDs to transmit the same PAM symbol. It is evident that APQ-SM exhibits an overall enhancement on the achievable SER performance, particularly for higher spectral efficiency values and for correlated MIMO setups. Thus, the proposed scheme is envisioned to provide an improvement on  the achievable throughput of SM-based schemes in VLC systems.  } 
\begin{figure}[ht]
\centering
\includegraphics[width=250pt]{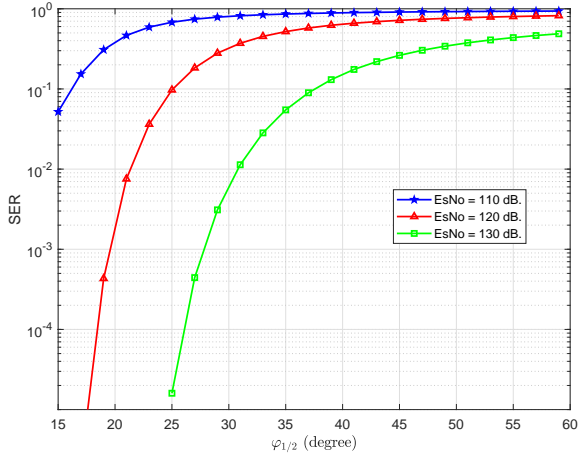}
\caption{ Effect of the semi-angle at half power on the performance of APQ-SM. }
\label{Fig7}
\end{figure}
Next, we quantify  the effect of varying the LED semi-angle on the achievable SER of the considered set up. This variation is considered for  different SNR values, as illustrated in Fig.\ref{Fig7}. To that end, it is observed that the best SER  performance  is  achieved  for  low $\varphi_{1/2}$ values.  Indeed, a small $\varphi_{1/2}$ leads to an increased Lambertian radiant intensity, which in turn improves  the quality of the corresponding communication channel, while reducing the incurred  channel correlation. 
On the contrary, a high $\varphi_{1/2}$ value degrades the quality of the wireless channel, which as a consequence increases the corresponding SER.

\begin{figure}[ht]
\centering
\includegraphics[width=250pt]{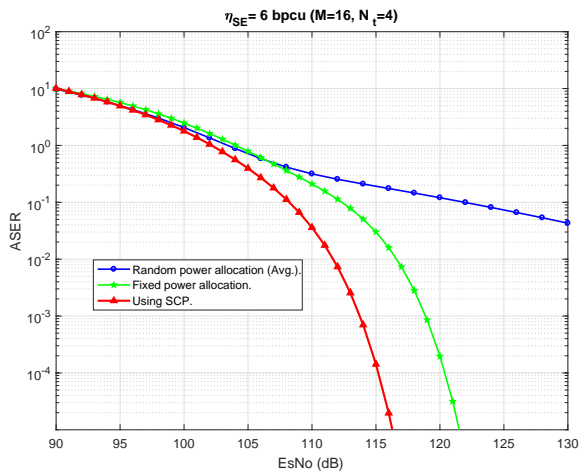}
\caption{ Comparison between the performance of the APQ-SM with random, fixed, and optimized power allocations for $\eta = 6\rm \, bpcu$. }
\label{Fig8}
\end{figure}

\begin{figure}[ht]
\centering
\includegraphics[width=250pt]{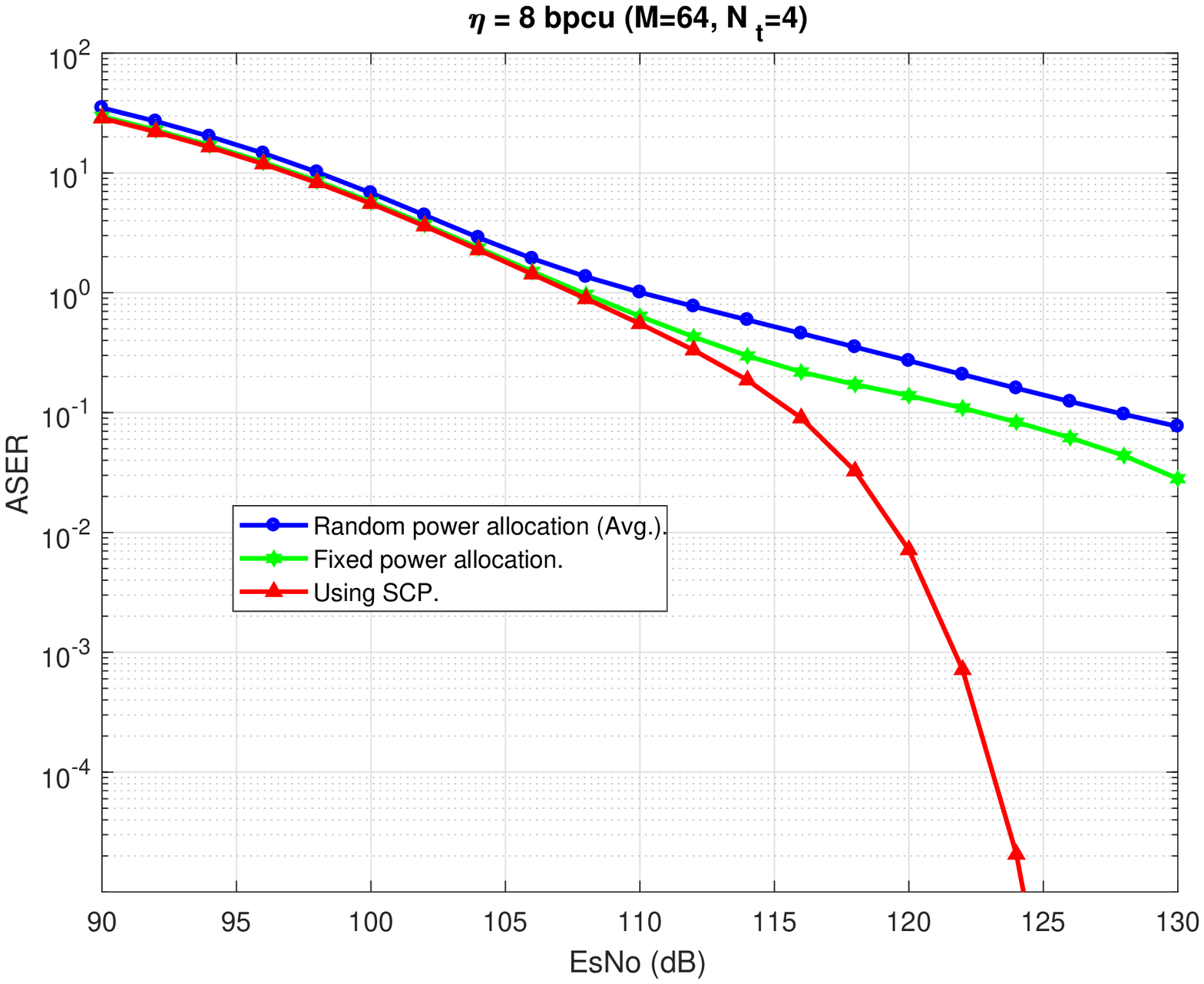}
\caption{ Comparison between the performance of the APQ-SM with random, fixed, and optimized power allocations for $\eta = 8\rm \,  bpcu$. }
\label{Fig9}
\end{figure}

\begin{figure}[ht]
\centering
\includegraphics[width=250pt]{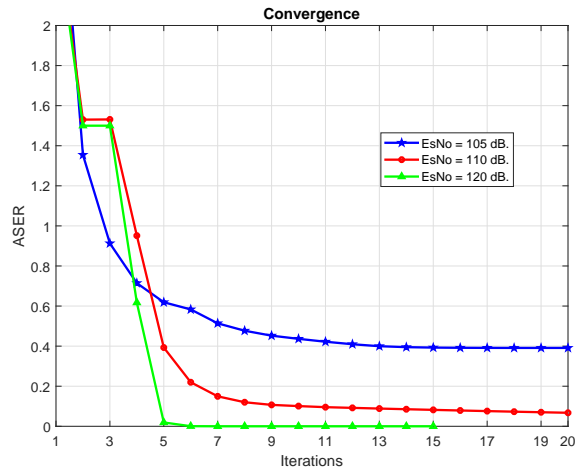}
\caption{ The convergence behaviour of Algorithm 1. }
\label{Fig10}
\end{figure}

Since the power allocation for each part of the APQ symbol is an essential factor that determines the
end to end  error rate  performance of the considered VLC system, we consider the results obtained by solving (P2). 
Based on this and without   loss of generality, we select the following indicative values for the  parameters in Algorithm 1: $\alpha_0=0.1,\; \alpha_1=0.9,\; \alpha_2=1,\; \alpha=1.5, \;\beta=2, \;\delta^{(0)}= 4,\;\epsilon=10^{-3},\; \text{and}\; N_{\rm max}=100$ \cite{Nocedal2006}.  
To this end, we compare the achievable ASER performance under random, fixed and SCP-optimized power allocation factors. Fig. \ref{Fig8} and Fig. \ref{Fig9} demonstrate the ASER for the indicative case of  $\eta=6 \rm \, bpcu$ and $\eta=8 \rm \, bpcu$, respectively. 
It becomes evident that  random power allocation exhibits  the worst achievable ASER performance. On the contrary, the optimum power allocation provides the best performance compared to the random and fixed power allocation approaches. Finally, we investigate in Fig. \ref{Fig10} the iteration convergence behavior of
Algorithm 1 in terms of the achieved ASER for different SNR values.
It is noticed that for higher SNR values, the algorithm converges to the  best performance after approximately 5 iterations. This is because  increasing the SNR values limits the optimization search space, which in turn  provides a faster result. 
In contrast, the algorithm converges after almost 13 iterations in the case of lower SNR values. 

\section{Conclusion}

This work proposed an effective spatial modulation based scheme, referred to as APQ-SM, which enhances  the performance of VLC systems. 
The proposed scheme transmits information in both the spatial and signal domains in order to increase the achievable spectral efficiency compared to related existing modulation schemes. In addition, the proposed scheme allows the transmission of higher order complex symbols by splitting the signal domain information into three different parts namely, amplitude, phase, and quadrant. Subsequently, each part was modulated using $M_i$-ary PAM modulation and then all parts were multiplexed in power domain and were transmitted from the active LED utilizing the entire available bandwidth. In this context, an explicit analytic expression was derived for the corresponding  ASER of the proposed scheme using both joint and proposed two-step detection methods. It was shown  that the proposed scheme is robust to the separation distance between the transmitting LEDs, whilst it achieves better ASER performance at low LEDs separations compared to traditional SM-PAM and MA-SM.  Furthermore, the offered analytic  and respective simulation
results revealed that the corresponding union bound can be  considered as a tight upper bound on the exact ASER, from moderate to high SNR values. Finally, in order to demonstrate the effect of the power allocation optimization, we formulated an optimization problem aiming to minimize the ASER. 
The  corresponding presented results showed that  optimized power allocation factors yield a significantly improved  ASER performance in comparison with random and fixed power counterpart  allocation schemes.  

\bibliographystyle{IEEEtran}
\bibliography{ref}

\begin{IEEEbiography}
[{\includegraphics[width=1in,height=1.25in,clip,keepaspectratio]{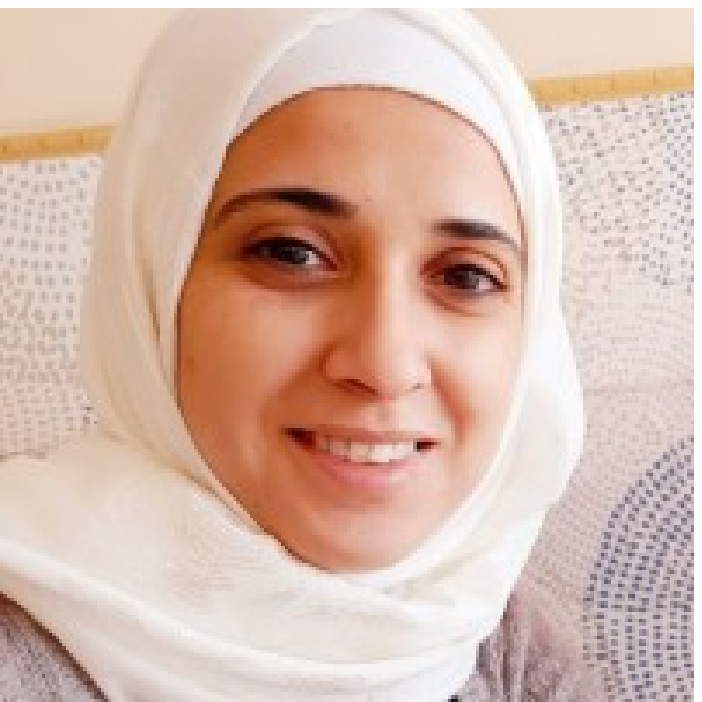}}]
{Shimaa Naser} received the M.Sc. degree in Electrical Engineering from the University of Science and Technology, Irbid, Jordan in 2015. Since 2018, she is pursuing her Ph.D. degree at the Department of Electrical and Computer Engineering at Khalifa University, Abu Dhabi, UAE. Her research interests  include  advanced  digital  signal  processing and modulation techniques for visible light communications, MIMO-based communiaction, and orthogonal/non-orthogonal  multiple  access.
\end{IEEEbiography}

\begin{IEEEbiography}
[{\includegraphics[width=1in,height=1.25in,clip,keepaspectratio]{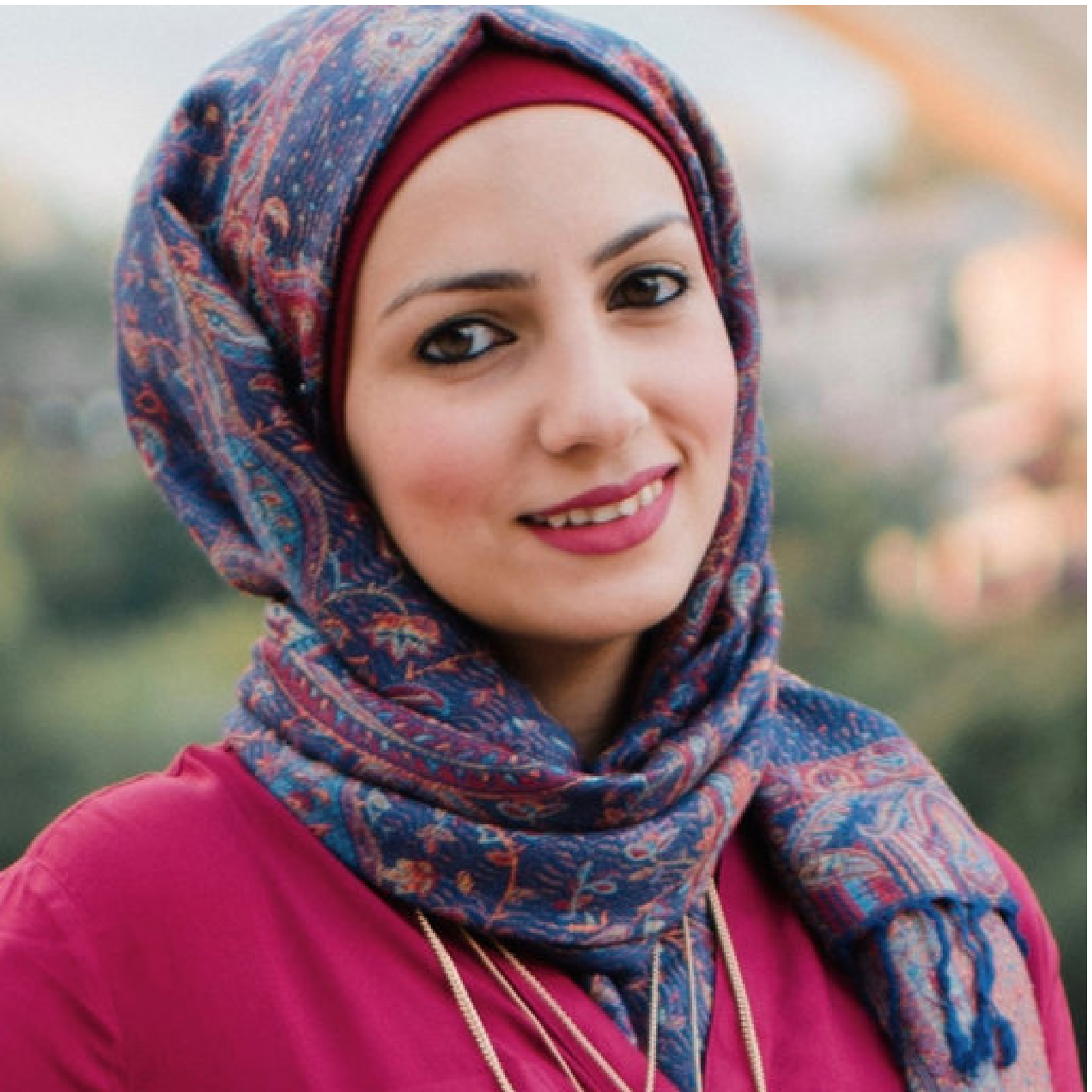}}]
{Lina Bariah} (S’13-M’19-SM'21) received the M.Sc. and Ph.D. degrees in communications engineering from Khalifa University, Abu Dhabi, UAE, in 2015 and 2018, respectively. She was a Visiting Researcher with the
Department of Systems and Computer Engineering, Carleton University, Ottawa, ON, Canada, in 2019. She is currently a Postdoctoral Fellow with the KU Center for Cyber-Physical Systems, Khalifa University, an affiliate research fellow, James Watt School of Engineering, University of Glasgow, UK, and an affiliate research fellow, University at Albany, SUNY, USA. Her research interests include advanced digital signal processing techniques for communications, machine learning, cooperative communications, non-orthogonal multiple access, cognitive radios, reconfigurable intelligent surfaces, aerial networks, and visible light communications. Dr. Bariah was a member of the technical program committee of a number of IEEE conferences, such as ICC and Globecom. She is currently an Associate Editor for the IEEE Open Journal of the Communications Society, an Editor for the IEEE Communications Letters, and an Area Editor for Physical Communication (Elsevier). She is a Guest Editor in RS Open Journal on Innovative Communication Technologies (RS-OJICT). She serves as a session chair, trackchair, and an active reviewer for numerous IEEE conferences and journals.

\end{IEEEbiography}

\begin{IEEEbiography}
[{\includegraphics[width=1in,height=1.25in,clip,keepaspectratio]{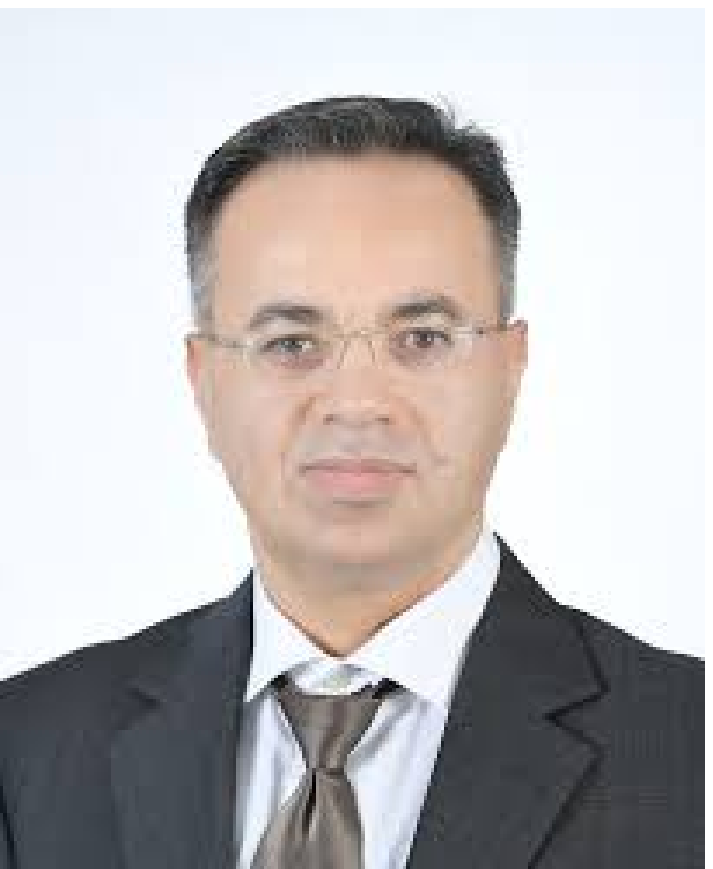}}]
{Sami Muhaidat} (S’01-M’08-SM’11) received the Ph.D. degree in electrical and computer engineering from the University of Waterloo, Waterloo, ON, Canada, in 2006. From 2007 to 2008, he was an NSERC Post-Doctoral Fellow with the Department of Electrical and Computer Engineering, University of Toronto, ON, Canada. From 2008 to 2012, he was Assistant Professor with the School of Engineering Science, Simon Fraser University, Burnaby, BC, Canada. He is
currently an Associate Professor with Khalifa University, Abu Dhabi, UAE, and a Visiting Professor with the Department of Electrical and Computer Engineering, University of Western
Ontario, London, ON, Canada. He is also a Visiting Reader with the Faculty of Engineering, University of Surrey, Guildford, U.K. 

Dr. Muhaidat currently serves as an Area Editor of the IEEE TRANSACTIONS ON COMMUNICATIONS, and he was previously a Senior Editor of the IEEE COMMUNICATIONS LETTERS and an Associate Editor of IEEE TRANSACTIONS ON COMMUNICATIONS, IEEE COMMUNICATIONS LETTERS and IEEE TRANSACTIONS ON VEHICULAR TECHNOLOGY. He was a recipient of several scholarships during his undergraduate and graduate studies and the winner of the 2006 NSERC PostDoctoral Fellowship Competition. Dr Muhaidat is a Senior Member IEEE.
\end{IEEEbiography}

\begin{IEEEbiography}
[{\includegraphics[width=1in,height=1.25in,clip,keepaspectratio]{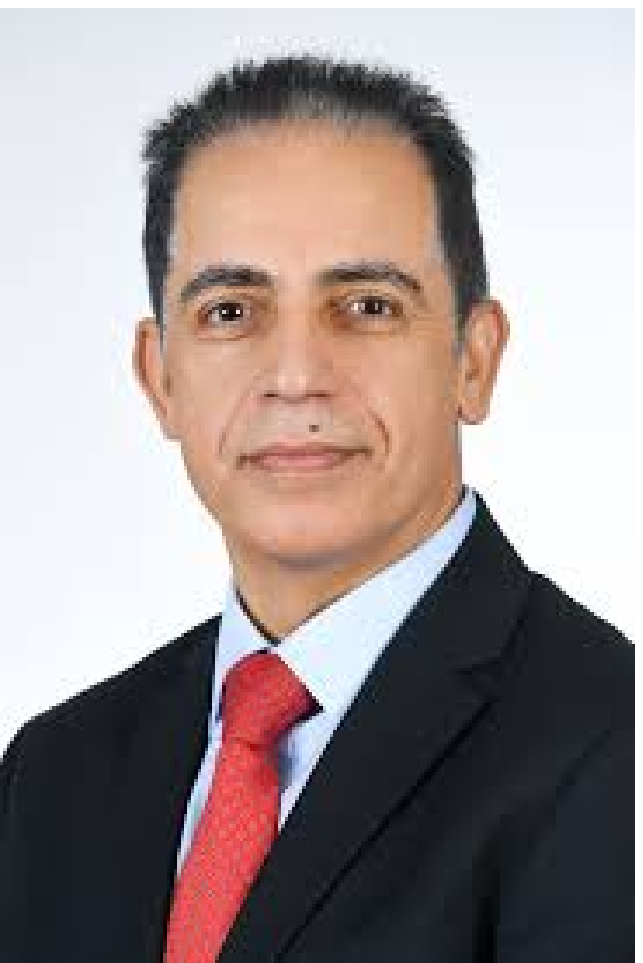}}]
{Mahmoud Al-Qutayri} (S'87, M'92, SM'02) is the Associate Dean for Graduate Studies – College of Engineering, and a Professor of Electrical and Computer Engineering at Khalifa University, UAE. He received the B.Eng., MSc and PhD degrees from Concordia University, Canada, University of Manchester, U.K., and the University of Bath, U.K., all in Electrical and Electronic Engineering in 1984, 1987, and 1992, respectively. Prior to joining Khalifa University, he was a Senior Lecturer at De Montfort University, UK. This was preceded by a Research Officer appointment at University of Bath, UK. He has published numerous technical papers in peer reviewed international journals and conferences. He coauthored a book as well as a number of book chapters. His main fields of research include embedded systems design, applications and security, design and test of mixed-signal integrated circuits, wireless sensor networks, and cognitive wireless networks. 

During his academic career, Dr. Al-Qutayri made many significant contributions to both undergraduate as well as graduate education. His professional service includes membership of the steering, organizing and technical program committees of many international conferences.
\end{IEEEbiography}

\begin{IEEEbiography}
[{\includegraphics[width=1in,height=1.25in,clip,keepaspectratio]{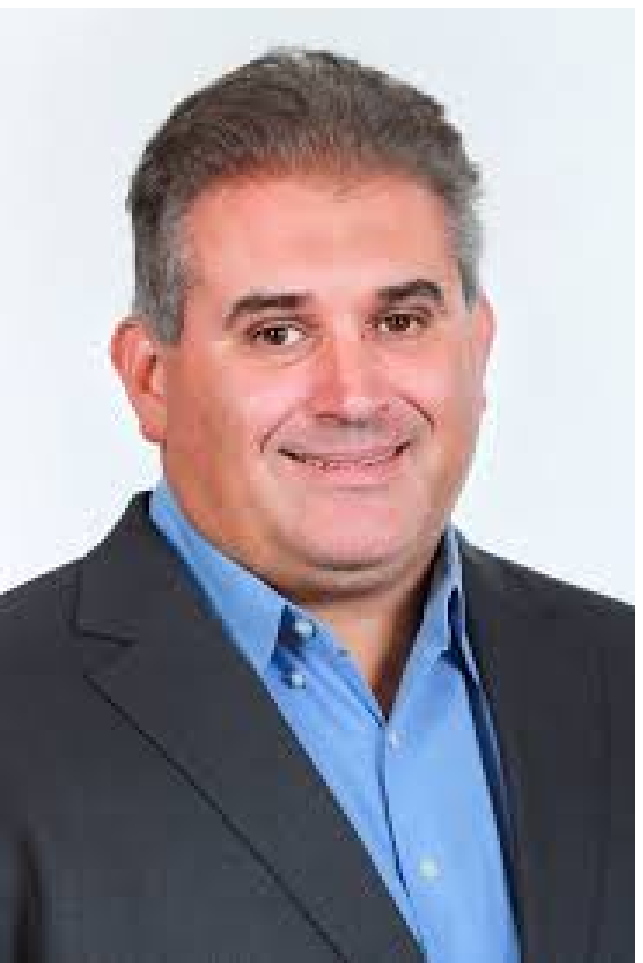}}]
{Paschalis C. Sofotasios} (S’07-M’12-SM’16) was born in Volos, Greece, in 1978. He received the M.Eng. degree from Newcastle University, U.K., in 2004, the M.Sc. degree from the University of Surrey, U.K., in 2006, and the Ph.D. degree from the University of Leeds, U.K., in 2011. He has held academic positions at the
University of Leeds, U.K., University of California at Los Angleles, CA, USA, Tampere University of Technology, Finland, Aristotle University of
Thessaloniki, Greece and Khalifa University of Science and Technology, UAE, where he currently serves as Associate Professor in the department of Electrical Engineering and Computer Science.
His M.Sc. studies were funded by a scholarship from UK-EPSRC and his Doctoral studies were sponsored by UK-EPSRC and Pace plc. His research interests are in the broad areas of digital and optical wireless communications as well as in topics relating to special functions and
statistics. 

Dr. Sofotasios serves as a regular reviewer for several international journals and has been a member of the technical program
committee of numerous IEEE conferences. He currently serves as an Editor for the IEEE COMMUNICATIONS LETTERS and he received the Exemplary Reviewer Award from the IEEE COMMUNICATIONS LETTERS
in 2012 and the IEEE TRANSACTIONS ON COMMUNICATIONS in 2015 and 2016. Dr. Sofotasios is a Senior Member IEEE and he received the Best Paper Award at ICUFN 2013. 
\end{IEEEbiography}

\end{document}